\documentclass[aps,prl,twocolumn,showpacs,superscriptaddress,groupedaddress]{revtex4-1}  
\usepackage{graphicx}  
\usepackage{amssymb,amsmath}   
\usepackage{color}
\usepackage{times}
\newcommand{\coloronline}{\textbf{Color online.~}}
\newcommand{\GammaN}{\Gamma}
\newcommand{\sectionhead}[1]{}

\newcommand{\eref}[1]{Eqn.~\ref{#1}}
\newcommand{\fref}[1]{Fig.~\ref{#1}}

\begin{document}
\title{Equilibrating temperature-like variables in jammed granular subsystems}
\author{James~G.\ Puckett and Karen~E.\ Daniels}
\affiliation{
Department of Physics, North Carolina State University, Raleigh, NC, USA 27695
}
\date{\today}

\begin{abstract}
Although jammed granular systems are athermal, several thermodynamic-like descriptions have been proposed which make quantitative predictions about the distribution of volume and stress within a system and provide a corresponding temperature-like variable. We perform experiments with an apparatus designed to generate a large number of independent, jammed, two-dimensional configurations. Each configuration consists of a single layer of photoelastic disks supported by a gentle layer of air. New configurations are generated by alternately dilating and re-compacting the system through a series of boundary displacements. Within each configuration, a bath of particles surrounds a smaller subsystem of particles with a different inter-particle friction coefficient than the bath. The use of photoelastic particles permits us to find all particle positions as well as the vector forces at each inter-particle contact. By comparing the temperature-like quantities in both systems, we find compactivity (conjugate to the volume) does not equilibrate between the systems, while the angoricity (conjugate to the stress) does.  Both independent components of the angoricity are linearly dependent on the hydrostatic pressure, in agreement with predictions of the stress ensemble.
\end{abstract}
\pacs{ 45.70.-n, 64.30.-t, 83.80.-Fg}
\maketitle

%
\sectionhead{Introduction}
Granular materials are a collection of discrete, athermal particles. In the absence of an external driving force, these materials relax into a mechanically stable jammed state and cannot move into another configuration since thermal fluctuations are negligible~\cite{Jaeger1996a}.  While these materials are therefore inherently non-equilibrium, preparing a configuration with a strict protocol nonetheless yields different microscopic states with the same, reproducible volume~\cite{Knight1995}.  Edwards proposed that the system volume (a conserved quantity) could be used to write a granular density of states, a corresponding entropy, and a temperature-like variable conjugate to the volume~\cite{Edwards1989}.  However, a complete granular statistical mechanics should describe the distribution of contact forces as well as the volumes. Subsequent theoretical advances have proposed that a stress-based ensemble~\cite{Ball2002, Goddard2004, Edwards2005, Henkes2007, Henkes2009, Blumenfeld2009, Lois2009, Tighe2011} is likely required for a full treatment. 

In the Edwards ensemble, the volume $V$ plays a role analogous to that of energy in equilibrium statistical mechanics. A granular temperature, dubbed the compactivity, is defined as $X \equiv (\partial S / \partial V)^{-1}$, and has been successfully measured in models~\cite{Srebro2003, Bowles2011},  simulations~\cite{Song2008, Briscoe2008}, and experiments~\cite{Nowak1998, Schroter2005, Lechenault2006, Ribiere2007, McNamara2009, Zhao2012a}. Similarly, the stress ensemble considers force and torque constraints on individual particles, and writes the density of states as a function of the stress-tensor $\widehat{\Sigma} = \sum \vec{r}_{ij} \vec{f}_{ij}$, where the $\vec{r}_{ij}$ are the vectors pointing from the center of each particle to its contacts, and $\vec{f}_{ij}$ is the corresponding contact force. The conjugate variable is then a tensorial temperature known as the angoricity, and is defined to be $\widehat{A} = (\partial S / \partial \widehat{\Sigma})^{-1}$. 

A minimal test of such temperature-like variables is to consider whether they obey the zeroth law of thermodynamics. In experiments and simulations, the compactivity~\cite{McNamara2009} has previously been shown to be equal in different parts of the same packing, and in different packings generated with the same particles under identical conditions. Simulations show this is also satisfied by the angoricity~\cite{Henkes2007, Henkes2009}. However, no test has been made of whether two dissimilar systems can equilibrate either $X$ or $\widehat{A}$. We provide such a test in a real granular system subject to isotropic compression, and find that while the compactivity fails this simple test, the angoricity equilibrates in a temperature-like way. 

\begin{figure}
\centering
\includegraphics[width=0.95\linewidth]{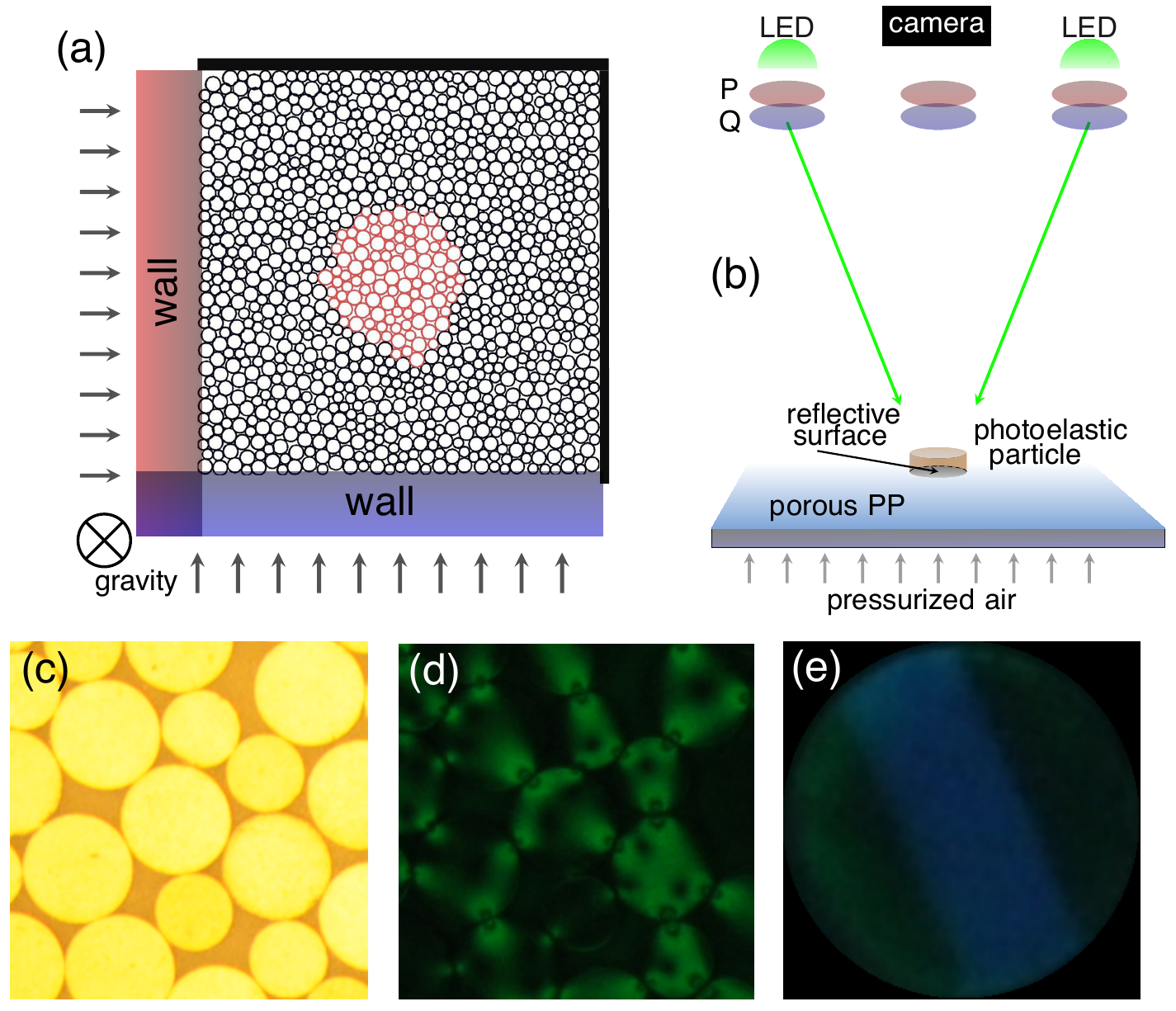}
\caption{\coloronline Schematic of apparatus showing (a) two walls bi-axially compressing an array of disk-shaped particles composed of an outer subsystem (black, high $\mu$) and an inner subsystem (red, low $\mu$) and (b) reflective photoelasticity on air-floated particles. Light shines from green LEDs through a linear polarizer (P) a wavelength-matched quarter wave plate (Q) before entering the photoelastic material. A mirrored surface on the bottom of each particle reflects light back through the particle. A second quarter-wave plate and linear polarizer are mounted on the camera to resolve the photoelasticity. Three images of each configuration are recorded: (c) unpolarized white light for locating particle positions, (d) polarized green light showing isochromatic fringes for calculating contact forces and (e) an ultraviolet light for identifying the low-$\mu$ particles.}
\label{fig:exp}
\end{figure}

Our experiments are conducted on a bi-disperse granular monolayer of photoelastic disks resting on a nearly frictionless surface provided by a thin layer of pressurized air. The assembly of particles is comprised of an inner subsystem and a larger bath which differ only in the inter-particle friction coefficient (see \fref{fig:exp}). Starting from a dilute state, the monolayer is bi-axially compressed by outer walls in a series of short steps. At some global volume fraction $\Phi$, the system jams and for all further steps the pressure on the system increases. Finally, the walls re-dilate to permit large scale rearrangements before the next series begins. By repeating this protocol many times, we generate an ensemble of configurations for which we record particle positions and calculate contact forces using methods similar to~\cite{Majmudar2006, Majmudar2007}.  With this information, we calculate the compactivity and angoricity for both the bath and the inner subsystem.

In the canonical volume ensemble~\cite{Edwards1989}, the probability of finding a system with volume $V$ and compactivity $X$ is proposed to be given by a Boltzmann-like distribution
\begin{equation}
{\cal{P}}(V) = \frac{\Omega(V)}{Z(X)} e^{-V/X}
\label{eqn:V}
\end{equation}
where the density of states is $\Omega(V)$ is defined for an ensemble of jammed configurations, and the partition function is $Z(X)$.  The stress ensemble similarly proposes a Boltzmann-like distribution 
\begin{equation}
{\cal{P}}(\widehat{\Sigma}) = \frac{\Omega(\widehat{\Sigma})}{Z(\widehat{A})} e^{-\mathrm{Tr}~( \widehat{\Sigma} / \widehat{A})}
\label{eqn:Gamma}
\end{equation}
for the stress-moment tensor $\widehat{\Sigma}$; the angoricity $\widehat{A}$ is therefore also a tensor. 

To calculate either $X$ or $\widehat{A}$, we use two methods: the method of overlapping histograms~\cite{Dean2003,Henkes2007,McNamara2009} and the fluctuation-dissipation theorem (FDT)~\cite{Nowak1998, Schroter2005, Zhao2012a}. The ratio of ${\cal{P}}(V)$ between two systems is exponential in $V$ and is given by
\begin{equation}
\frac{{\cal{P}}_1(V)}{{\cal{P}}_2(V)} = \frac{Z(X_2)}{Z(X_1)}~~  e^{\left( \frac{1}{X_2} - \frac{1}{X_1} \right) V}.
\label{eqn:volume}
\end{equation}
By taking the logarithm of this ratio, one obtains a term linear in $V$, where the coefficient is the difference in the inverse temperatures.  This method determines $1/X$ up to an additive constant: $1/X \rightarrow 1/X +C_X$.
The FDT method also provides a relative measurement. Using the measured variance $\langle \delta V^2 \rangle$ of ${\cal{P}}(V)$, we compute 
\begin{equation}
\frac{1}{X_1}-\frac{1}{X_2} = \int_{V_1}^{V_2} \frac{d V}{\langle \delta V^2 \rangle}\quad
\label{eqn:fdt}
\end{equation}
to obtain values of $X$, also up to a constant. The calculation of $\widehat{A}$ utilizes equations analogous to \eref{eqn:volume} and \eref{eqn:fdt}; the tensorial aspects will be discussed in more detail below. Each of these methods is used separately on  both the subsystem and the bath, in order to test for equilibration.

%

\sectionhead{Experiment}
Our experimental apparatus is shown to scale in \fref{fig:exp}. The granular monolayer consists of $1004$ bi-disperse photoelastic (Vishay PhotoStress PSM-4) disks with a thickness $\approx 3.1$~mm and diameters $d_S = 11.0$~mm and $d_L = 15.4$~mm, in equal concentrations.  The particles are supported on a thin layer of air provided by a steady flow of pressurized air through a porous polypropylene sheet with a nominal pore size of $120~\mu$m.  This minimizes the effect of friction between the particles and the surface, but does not  otherwise cause significant dynamics once the system is jammed. The sheet is leveled (particles do not drift to one side) and flat (particles do not cluster). The system consists of an outer bath $N_{B} = 904$ and an inner subsystem $N_{S} = 100$.  Particles in the bath have a friction coefficient $\mu_B \approx 0.8$, while particles in the inner subsystem are wrapped with a thin layer of PTFE tape with a $\mu_S < 0.1$.

Images of the particle positions, photoelastic images for measuring vector contact forces, and identification of the subsystem particles are recorded with three separate images captured by a single CCD camera located above the apparatus (see \fref{fig:exp}b). Particle positions are identified using a white light image (see \fref{fig:exp}c), from which the centers are detected  with an accuracy  of $\approx  0.01 d_S$ using a Hough transform. 
The photoelastic images (see \fref{fig:exp}d) are captured using reflective photoelasticity, in which the silvered back side of each particle reflects polarized light back to the camera. Photoelasticity allows for the numerical determination of the normal and tangential forces at each contact point, as required to measure $\widehat{\Sigma}$. Similar to the methods pioneered by \cite{Majmudar2006,Majmudar2007}, we minimize the error between the observed and fitted image of the particle using a non-linear least-squares optimization. Details and source code are available for download at \cite{peDiscSolve}.
The third image is taken using black-light illumination to identify the subsystem particles, which are tagged with ultraviolet-sensitive ink (see \fref{fig:exp}e). The subsystem comprises all low-$\mu$ particles which are Vorono\"i neighbors with at least one other particle in the subsystem.

The particles are confined within a square region (maximally $50\times50$~cm) imposed by two stationary walls positioned by stepper motors, as shown in \fref{fig:exp}a. The system is initially in a dilute, un-jammed state, with the global volume fraction  $\Phi \lesssim 0.6$. The two walls bi-axially compress the system by a series of small steps of constant size ($\Delta \Phi = 0.0009$, equivalently $\Delta x = 0.3$~mm or $0.02~r_L$).  With each step of the wall, the three images are recorded, and data is collected over a series of volumes corresponding to $0.775 < \Phi < 0.805$, giving $30$ different volumes for each compression cycle.
Steps continue until the gradient squared of the force image~\cite{Howell1999} indicates a pressure threshold has been reached; this reduces the risk of particles buckling out of plane. The walls then re-dilate to the dilute state, and the particles are then rearranged while maintaining subsystem continuity; this protocol is repeated $100$ times. 

\begin{figure}
\centering
\includegraphics[width=0.95\columnwidth]{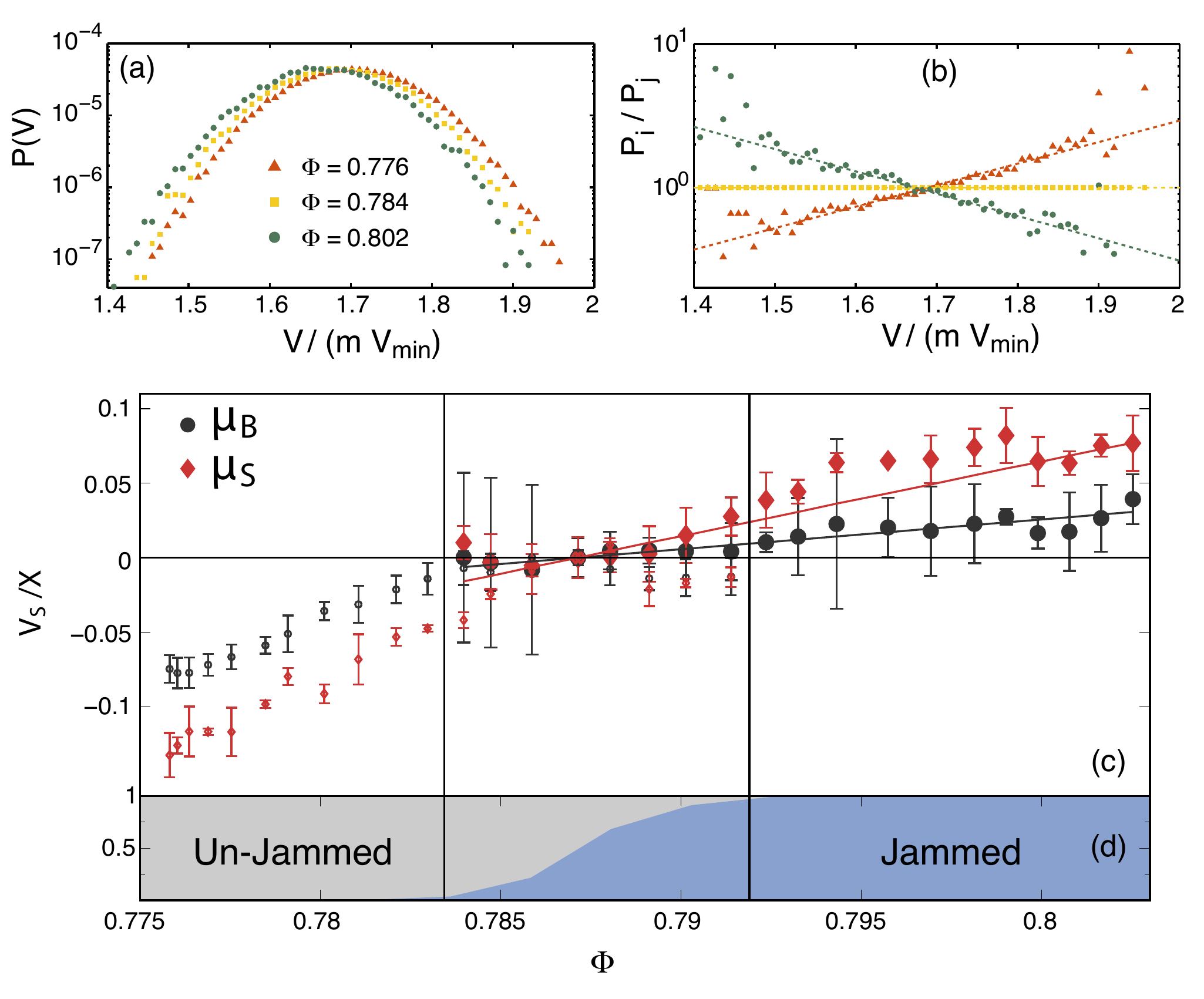}
\caption{\coloronline (a) Volume histograms, ${\cal P}(V)$, for $\Phi = 0.776$ ($\blacktriangle$), $0.784$ ($\blacksquare$), and $0.802$ ($\bullet$) with $m=48$. (b) A semi-logarithmic plot of the ratio each histogram with respect to the $\Phi = 0.784$ distribution, i.e. ${\cal P}_i(V)/{\cal P}_{i=2}(V)$. (c) The inverse compactivity given by \eref{eqn:volume} plotted as a function of the inverse volume fraction where $\mu_B$ are shown as black \textbullet ~and $\mu_S$ are red $\blacklozenge$. Large/small symbols denote jammed/un-jammed configurations, respectively. Errorbars shown are  uncertainties in ${\cal P}(V)$ and propagated through the calculation.  The inverse compactivity given by the FDT method (\eref{eqn:fdt}), is shown with the solid line for comparison.  (d) The ratio of number of jammed/un-jammed configurations recorded at each $\Phi$. }
\label{fig:volume}
\end{figure}

During the compression phase of each quasi-static cycle, we observe the percolation of force chains at a value $\Phi_{perc}$. As the system is further compressed beyond this point, the contact forces grow in strength and the average number of contacts per particle increases. For the set of $100$ cycles, this threshold occurs over a range $0.782 < \Phi_{perc} < 0.792$, where the width of the distribution is indicative of finite size effects \cite{OHern2003,Shen2012b}. The ratio of un-jammed to jammed systems at a given $\Phi$ is shown in \fref{fig:volume}d. We define random loose packing as $\Phi_{RLP} = \langle \Phi_{perc} \rangle \approx 0.787$ as the center of this distribution.

%

\sectionhead{Results Volume}

We calculate the distribution of local volumes  ${\cal P}(V_m)$ over clusters of size $m$, using the sum of individual radical Vorono\"i volumes obtained from the Voro++ software~\cite{Rycroft2006}. Each cluster is defined as the $m-1$ nearest neighbors surrounding a central particle. For $m = 1$, ${\cal P}(V_m)$ has two distinct peaks which correspond to small and large particles \cite{Lechenault2006}. With increasing cluster size, the bimodal aspect of ${\cal P}(V_m)$ disappears, but even for large cluster sizes ($m>100$), the distribution remains asymmetric and non-Gaussian~\cite{Lechenault2006}.  In \fref{fig:volume}a, we show ${\cal P}(V_m)$ for three values of $\Phi$ with $m = 48$; the value of $m$ is large enough so that ${\cal P}(V_m)$ does not show any features arising from bi-dispersity.

In \fref{fig:volume}b, we show the ratio ${\cal P}_i(V)/{\cal P}_j(V)$ where the reference system $j$ is $\Phi = 0.784$.  In practice this can be done with any two systems so long as there is sufficient overlap between their histograms. As the ratio of ${\cal P}_i(V) / {\cal P}_j(V)$ is well-approximated by an exponential in $V$, the compactivity can be calculated using \eref{eqn:volume}. The inverse compactivity, $1/X$, is also calculated using FDT using \eref{eqn:fdt}, where the integrand is approximated using a third order polynomial. Each method determines $1/X$ only up to an additive constant, which is adjusted so that $X_{RLP} = \infty$. In \fref{fig:volume}c, the inverse compactivity is shown for both the bath ($1/X_B$) and the subsystem ($1/X_S$).  We find good agreement between $X(\Phi)$ given by the overlapping histogram method and by the fluctuation dissipation theorem.  In addition, for $4 < m < 50$, we observe $X$ to be approximately independent of $m$. However, we find that the compactivity of the bath is not equal to that of the subsystem ($X_B(\Phi) \neq X_S(\Phi)$), even considering adjustments of the additive constant. This represents a failure of the zeroth law for $X$.

We can take further advantage of the accessibility of both jammed and un-jammed states within in the center of the range of explored $\Phi$. While the Edwards ensemble is not defined for un-jammed systems, we can nonetheless carry out the histogram analysis as performed on the jammed systems. In this regime, we find that the ${\cal P}(V_m)$ histograms cannot distinguish between the jammed and un-jammed states. Furthermore, the measured values of $X$  decrease continuously from above $\Phi_{RLP}$ to below; this is an undesirable characteristic.


%

\sectionhead{Results stress}

The stress ensemble also provides a Boltzmann-like distribution in the stress, as given in \eref{eqn:Gamma}. In the case of frictionless grains, the angoricity $\widehat{A}$ is a scalar due the off-diagonal components in $\widehat{\Sigma}$ being zero.  In any real granular system, friction is present and a shear-free state is not readily obtained. Therefore, $\widehat{\Sigma}$ is a symmetric tensor with non-zero off-diagonal components and can be reduced to two independent components related to the pressure and shear stress. The pressure angoricity $A_p$ and the shear angoricity $A_\tau$ are conjugate to $\sigma_p = (\sigma_1+\sigma_2)/2$ and the $\sigma_\tau = (\sigma_1-\sigma_2)/2$, respectively~\cite{Henkes2009}, where $\sigma_{1,2}$ are the principal stresses. The average hydrostatic pressure per particle in the system is  given by $\displaystyle \GammaN = \mathrm{Tr}~\widehat{\Sigma} / N$.  Both $A_p$ and $A_\tau$ are obtained using the method of overlapping histograms (analogous to \eref{eqn:volume}) and the FDT (analogous to \eref{eqn:fdt}). With each method, $A$ is calculated up to an additive constant so that $1/A \rightarrow 1/A + C_A$, where $C_A$ satisfies $A \rightarrow \infty$ as $\GammaN \rightarrow \infty$.

\begin{figure}
\centering
\includegraphics[width=0.95\columnwidth]{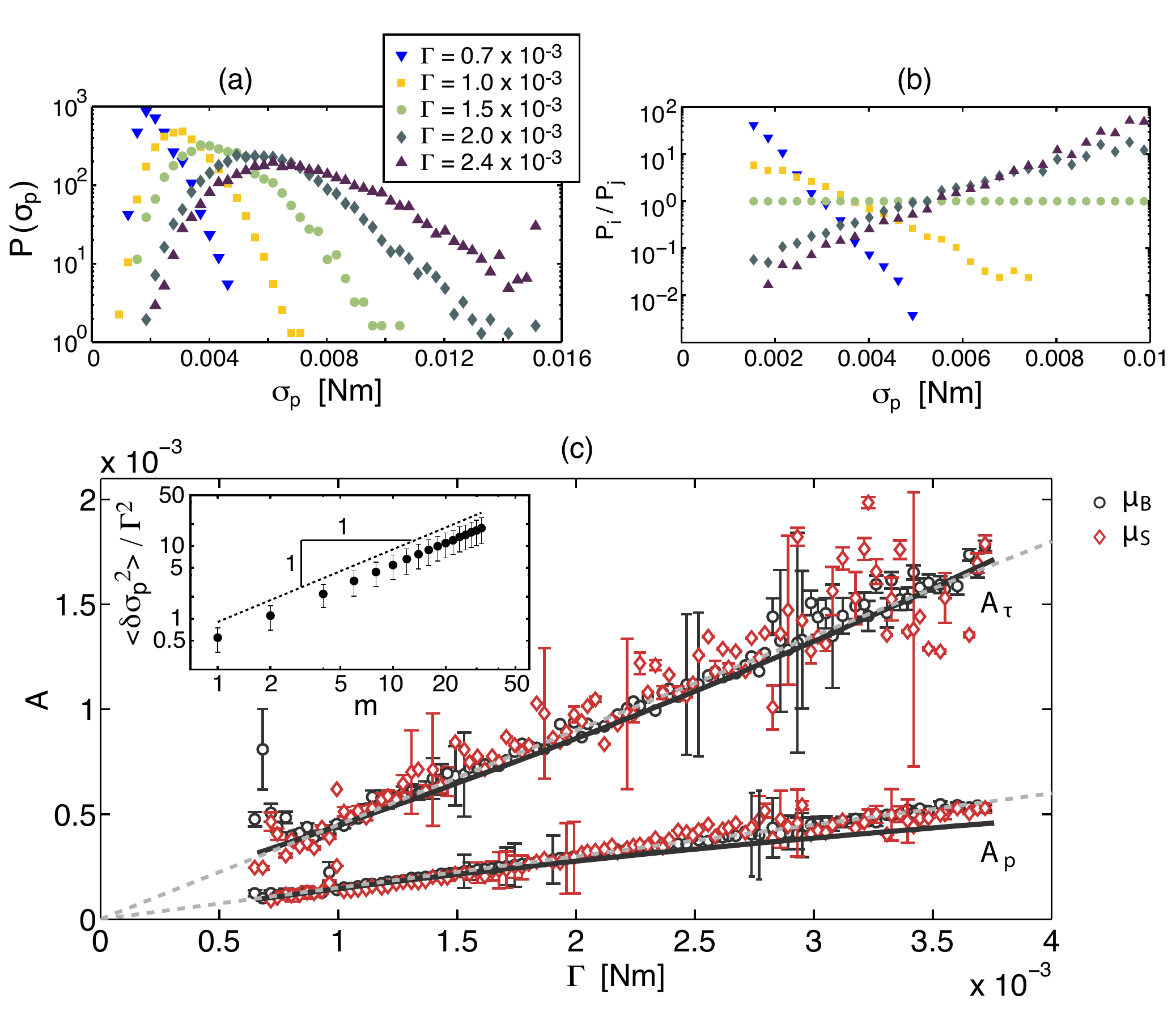}
\caption{\coloronline (a) Distribution of $\sigma_p$ where $m=8$ and $\GammaN = 0.0007$ ($\blacktriangledown$), $0.0010$ ($\blacksquare$), $0.0015$ ($\bullet$), and $0.0024$~Nm ($\blacktriangle$). A semi-logarithmic plot of the (b) ratio ${\cal P}_i(\sigma_p)/{\cal P}_j(\sigma_p)$ where the reference system $j$ is $\GammaN = 0.0015$~Nm. The pressure angoricity $A_P$ and shear angoricity $A_\tau$ are shown as a function of $\GammaN$ where the results using overlapping histograms for $\mu_B$ and $\mu_S$ are shown as black $\circ$ and are red $\diamondsuit$, respectively.  The solid line is the angoricity calculated using FDT. The gray dashed lines provide a visual reference of the slopes $0.15$ and $0.45$, respectively.  Inset: The scaled variance $\langle \delta\sigma_p^2 \rangle$  of the ${\cal P}_j(\sigma_{p})$ distribution, as a function of the cluster size $m$.}
\label{fig:stress}
\end{figure}

In \fref{fig:stress}a, the local distribution of pressure ${\cal P}(\sigma_{p})$ is shown for $m = 8$ on configurations over a range $0.0006 < \GammaN < 0.0025$~Nm.  
The ratio ${\cal P}_i(\sigma_{p})/{\cal P}_j(\sigma_{p})$ is exponential in $\sigma_p$ (see \fref{fig:stress}b, similar results for $\sigma_\tau$ not shown), as required by the stress ensemble analogue of \eref{eqn:volume}. In addition, we observe that the variance of $\sigma_p$ is proportional to $m$, which is consistent with $S$ being an extensive entropy (see  \fref{fig:stress}c). 
We are therefore able to measure both the pressure angoricity $A_p$ the shear angoricity $A_\tau$ using their corresponding distributions, shown in \fref{fig:stress}c as a function of $\GammaN$. 
We find that $A_{p,\tau}$  are independent of $m$ for $m > 3$, as also observed in simulations~\cite{Henkes2007, Henkes2009}, and that values obtained from the histogram method (points) and the FDT method (solid line) are in approximate agreement. Finally, we find that for either the shear or compressional angoricity, the values measured in the bath and in the subsystem are equivalent, signifying the angoricity is equilibrating between the subsystems.  

Nonetheless, the values of $A_\tau$  and $A_p$ do not match each other, with the shear angoricity growing faster as a function of  $\GammaN$. We find the angoricity is given by $A = b~\GammaN$ for both pressure angoricity and shear angoricity, where $b_p = 0.153 \pm 0.004$ and $b_\tau = 0.450 \pm 0.020$, respectively.  For a two-dimensional frictionless shear-free system, the stress ensemble predicts $b_p = 0.5$ at the isostatic point~\cite{Henkes2007,Henkes2009}.  Above the isostatic point, the stress ensemble predicts $b_p$ to be a function of the average contact number.  The disagreement between the \emph{frictional} and \emph{frictionless} values of $b_p$ implies friction significantly affects the density of states.

%
\sectionhead{Discussion}

We have measured both compactivity $X$ (conjugate to volume in the Edwards ensemble), and angoricity $\widehat{A}$ (conjugate to the stress tensor in the stress ensemble), in a laboratory granular system using particle-scale characterizations. While we found that while the value of $X$ calculated using the overlapping histogram method was consistent with the value found using the fluctuation-dissipation theorem, it failed to equilibrate between non-identical systems, making it a poor state variable. A similar failure is likely behind previous measurements by \citet{Schroter2005}, in which two granular materials with different frictional properties, prepared using the same protocol, were found to have different globally-measured values of $X$. In contrast, we observed that the temperature-like variable $\widehat{A}$ does successfully equilibrate between a subsystem and bath with dissimilar inter-particle friction coefficients, as would be required in order to have a valid zeroth law. Moreover, we find agreement with the prediction that angoricity should scale linearly the hydrostatic pressure~\cite{Henkes2009}. These successes make angoricity a promising state variable for frictional granular systems. 

One downside to using angoricity as a state variable, particularly in experiments, is that its calculation requires the determination of both normal and tangential forces. While there has been a long history of measuring normal forces at the boundaries of granular systems~\cite{Mueth1998, Lovoll1999, Blair2001, Makse2000, Corwin2005}, particle-scale measurements have seen more limited development. Outside of photoelastic particles such as those used here, measurements typically exist only for normal forces, whether the systems are frictional (tangential forces are neglected)~\cite{Mukhopadhyay2011, Saadatfar2012} or frictionless~\cite{Brujic2003,Zhou2006a, Desmond2012}.

It is possible to understand the success of the stress ensemble over the Edwards (volume) ensemble by considering the underlying physics behind the conserved quantities in each. Under Newton's third law, forces and torques must be strictly balanced at each force contact, while volume is merely constrained globally. As a result, our subsystem differed from the bath not only in the measured $X$, but more conventionally in the mean local volume fraction.

In fact, the full canonical Edwards ensemble~\cite{Edwards2005} unifies the volume and stress ensembles, where the density of states depends on both $V$ and $\widehat{\Sigma}$, and it has recently been argued~\cite{Wang2012d,Blumenfeld2012} that the two should not be considered separately.  The classic phenomenon of Reynolds dilatancy~\cite{Reynolds1885} under which shear induces a bulk expansion similarly suggests that such a coupling is important. Nonetheless, we observed here that angoricity can be independently equilibrated, and future experiments should more fully investigate the relationship between ensembles, the relative importance of shear and compression, and the role of friction on the density of states. 



\paragraph{Acknowledgements:}
The authors are grateful for financial support under from the National Science Foundation (DMR-0644743), and for illuminating discussions with Dapeng Bi, Bulbul Chakraborty, Silke Henkes, Brian Tighe, Matthias Schr\"oter, and Song-Chuan Zhao.

%


\begin{thebibliography}{42}%
\makeatletter
\providecommand \@ifxundefined [1]{%
 \@ifx{#1\undefined}
}%
\providecommand \@ifnum [1]{%
 \ifnum #1\expandafter \@firstoftwo
 \else \expandafter \@secondoftwo
 \fi
}%
\providecommand \@ifx [1]{%
 \ifx #1\expandafter \@firstoftwo
 \else \expandafter \@secondoftwo
 \fi
}%
\providecommand \natexlab [1]{#1}%
\providecommand \enquote  [1]{``#1''}%
\providecommand \bibnamefont  [1]{#1}%
\providecommand \bibfnamefont [1]{#1}%
\providecommand \citenamefont [1]{#1}%
\providecommand \href@noop [0]{\@secondoftwo}%
\providecommand \href [0]{\begingroup \@sanitize@url \@href}%
\providecommand \@href[1]{\@@startlink{#1}\@@href}%
\providecommand \@@href[1]{\endgroup#1\@@endlink}%
\providecommand \@sanitize@url [0]{\catcode `\\12\catcode `\$12\catcode
  `\&12\catcode `\#12\catcode `\^12\catcode `\_12\catcode `\%12\relax}%
\providecommand \@@startlink[1]{}%
\providecommand \@@endlink[0]{}%
\providecommand \url  [0]{\begingroup\@sanitize@url \@url }%
\providecommand \@url [1]{\endgroup\@href {#1}{\urlprefix }}%
\providecommand \urlprefix  [0]{URL }%
\providecommand \Eprint [0]{\href }%
\@ifxundefined \urlstyle {%
  \providecommand \doi  [0]{\begingroup \@sanitize@url \@doi}%
  \providecommand \@doi [1]{\endgroup \@@startlink {\doibase
  #1}doi:\discretionary {}{}{}#1\@@endlink }%
}{%
  \providecommand \doi  [0]{doi:\discretionary{}{}{}\begingroup
  \urlstyle{rm}\Url }%
}%
\providecommand \doibase [0]{http://dx.doi.org/}%
\providecommand \Doi [0]{\begingroup \@sanitize@url \@Doi }%
\providecommand \@Doi  [1]{\endgroup\@@startlink{\doibase#1}\@@Doi}%
\providecommand \@@Doi [1]{#1\@@endlink}%
\providecommand \selectlanguage [0]{\@gobble}%
\providecommand \bibinfo  [0]{\@secondoftwo}%
\providecommand \bibfield  [0]{\@secondoftwo}%
\providecommand \translation [1]{[#1]}%
\providecommand \BibitemOpen [0]{}%
\providecommand \bibitemStop [0]{}%
\providecommand \bibitemNoStop [0]{.\EOS\space}%
\providecommand \EOS [0]{\spacefactor3000\relax}%
\providecommand \BibitemShut  [1]{\csname bibitem#1\endcsname}%
\bibitem [{\citenamefont {Jaeger}\ \emph {et~al.}(1996)\citenamefont {Jaeger},
  \citenamefont {Nagel},\ and\ \citenamefont {Behringer}}]{Jaeger1996a}%
  \BibitemOpen
  \bibfield  {author} {\bibinfo {author} {\bibfnamefont {H.}~\bibnamefont
  {Jaeger}}, \bibinfo {author} {\bibfnamefont {S.~R.}\ \bibnamefont {Nagel}}, \
  and\ \bibinfo {author} {\bibfnamefont {R.~P.}\ \bibnamefont {Behringer}},\
  }\Doi {10.1103/RevModPhys.68.1259} {\bibfield  {journal} {\bibinfo  {journal}
  {Reviews of Modern Physics},\ }\textbf {\bibinfo {volume} {68}},\ \bibinfo
  {pages} {1259} (\bibinfo {year} {1996})}\BibitemShut {NoStop}%
\bibitem [{\citenamefont {Knight}\ \emph {et~al.}(1995)\citenamefont {Knight},
  \citenamefont {Fandrich}, \citenamefont {Lau}, \citenamefont {Jaeger},\ and\
  \citenamefont {Nagel}}]{Knight1995}%
  \BibitemOpen
  \bibfield  {author} {\bibinfo {author} {\bibfnamefont {J.~B.}\ \bibnamefont
  {Knight}}, \bibinfo {author} {\bibfnamefont {C.~G.}\ \bibnamefont
  {Fandrich}}, \bibinfo {author} {\bibfnamefont {C.~N.}\ \bibnamefont {Lau}},
  \bibinfo {author} {\bibfnamefont {H.~M.}\ \bibnamefont {Jaeger}}, \ and\
  \bibinfo {author} {\bibfnamefont {S.~R.}\ \bibnamefont {Nagel}},\ }\href
  {http://pre.aps.org/abstract/PRE/v51/i5/p3957\_1} {\bibfield  {journal}
  {\bibinfo  {journal} {Physical Review E},\ }\textbf {\bibinfo {volume}
  {51}},\ \bibinfo {pages} {3957} (\bibinfo {year} {1995})}\BibitemShut
  {NoStop}%
\bibitem [{\citenamefont {Edwards}\ and\ \citenamefont
  {Oakeshott}(1989)}]{Edwards1989}%
  \BibitemOpen
  \bibfield  {author} {\bibinfo {author} {\bibfnamefont {S.~F.}\ \bibnamefont
  {Edwards}}\ and\ \bibinfo {author} {\bibfnamefont {R.~B.~S.}\ \bibnamefont
  {Oakeshott}},\ }\Doi {10.1016/0378-4371(89)90034-4} {\bibfield  {journal}
  {\bibinfo  {journal} {Physica A},\ }\textbf {\bibinfo {volume} {157}},\
  \bibinfo {pages} {1080} (\bibinfo {year} {1989})}\BibitemShut {NoStop}%
\bibitem [{\citenamefont {Ball}\ and\ \citenamefont
  {Blumenfeld}(2002)}]{Ball2002}%
  \BibitemOpen
  \bibfield  {author} {\bibinfo {author} {\bibfnamefont {R.~C.}\ \bibnamefont
  {Ball}}\ and\ \bibinfo {author} {\bibfnamefont {R.}~\bibnamefont
  {Blumenfeld}},\ }\Doi {10.1103/PhysRevLett.88.115505} {\bibfield  {journal}
  {\bibinfo  {journal} {Physical Review Letters},\ }\textbf {\bibinfo {volume}
  {88}},\ \bibinfo {pages} {115505} (\bibinfo {year} {2002})}\BibitemShut
  {NoStop}%
\bibitem [{\citenamefont {Goddard}(2004)}]{Goddard2004}%
  \BibitemOpen
  \bibfield  {author} {\bibinfo {author} {\bibfnamefont {J.}~\bibnamefont
  {Goddard}},\ }\Doi {10.1016/j.ijsolstr.2004.05.049} {\bibfield  {journal}
  {\bibinfo  {journal} {International Journal of Solids and Structures},\
  }\textbf {\bibinfo {volume} {41}},\ \bibinfo {pages} {5851} (\bibinfo {year}
  {2004})}\BibitemShut {NoStop}%
\bibitem [{\citenamefont {Edwards}(2005)}]{Edwards2005}%
  \BibitemOpen
  \bibfield  {author} {\bibinfo {author} {\bibfnamefont {S.~F.}\ \bibnamefont
  {Edwards}},\ }in\ \href@noop {} {\emph {\bibinfo {booktitle} {Powders and
  Grains 2005}}},\ Vol.~\bibinfo {volume} {1},\ \bibinfo {editor} {edited by\
  \bibinfo {editor} {\bibfnamefont {R.}~\bibnamefont {Garcia-Rojo}}, \bibinfo
  {editor} {\bibfnamefont {H.~J.}\ \bibnamefont {Herrmann}}, \ and\ \bibinfo
  {editor} {\bibfnamefont {S.}~\bibnamefont {McNamara}}}\ (\bibinfo {year}
  {2005})\ p.~\bibinfo {pages} {3}\BibitemShut {NoStop}%
\bibitem [{\citenamefont {Henkes}\ \emph {et~al.}(2007)\citenamefont {Henkes},
  \citenamefont {O'Hern},\ and\ \citenamefont {Chakraborty}}]{Henkes2007}%
  \BibitemOpen
  \bibfield  {author} {\bibinfo {author} {\bibfnamefont {S.}~\bibnamefont
  {Henkes}}, \bibinfo {author} {\bibfnamefont {C.~S.}\ \bibnamefont {O'Hern}},
  \ and\ \bibinfo {author} {\bibfnamefont {B.}~\bibnamefont {Chakraborty}},\
  }\Doi {10.1103/PhysRevLett.99.038002} {\bibfield  {journal} {\bibinfo
  {journal} {Physical Review Letters},\ }\textbf {\bibinfo {volume} {99}},\
  \bibinfo {pages} {038002} (\bibinfo {year} {2007})}\BibitemShut {NoStop}%
\bibitem [{\citenamefont {Henkes}\ and\ \citenamefont
  {Chakraborty}(2009)}]{Henkes2009}%
  \BibitemOpen
  \bibfield  {author} {\bibinfo {author} {\bibfnamefont {S.}~\bibnamefont
  {Henkes}}\ and\ \bibinfo {author} {\bibfnamefont {B.}~\bibnamefont
  {Chakraborty}},\ }\Doi {10.1103/PhysRevE.79.061301} {\bibfield  {journal}
  {\bibinfo  {journal} {Physical Review E},\ }\textbf {\bibinfo {volume}
  {79}},\ \bibinfo {pages} {061301} (\bibinfo {year} {2009})}\BibitemShut {NoStop}%
\bibitem [{\citenamefont {Blumenfeld}\ and\ \citenamefont
  {Edwards}(2009)}]{Blumenfeld2009}%
  \BibitemOpen
  \bibfield  {author} {\bibinfo {author} {\bibfnamefont {R.}~\bibnamefont
  {Blumenfeld}}\ and\ \bibinfo {author} {\bibfnamefont {S.~F.}\ \bibnamefont
  {Edwards}},\ }\Doi {10.1021/jp809768y} {\bibfield  {journal} {\bibinfo
  {journal} {Journal of Physical Chemistry B},\ }\textbf {\bibinfo {volume}
  {113}},\ \bibinfo {pages} {3981} (\bibinfo {year} {2009})}\BibitemShut
  {NoStop}%
\bibitem [{\citenamefont {Lois}\ \emph {et~al.}(2009)\citenamefont {Lois},
  \citenamefont {Zhang}, \citenamefont {Majmudar}, \citenamefont {Henkes},
  \citenamefont {Chakraborty}, \citenamefont {O'Hern},\ and\ \citenamefont
  {Behringer}}]{Lois2009}%
  \BibitemOpen
  \bibfield  {author} {\bibinfo {author} {\bibfnamefont {G.}~\bibnamefont
  {Lois}}, \bibinfo {author} {\bibfnamefont {J.}~\bibnamefont {Zhang}},
  \bibinfo {author} {\bibfnamefont {T.~S.}\ \bibnamefont {Majmudar}}, \bibinfo
  {author} {\bibfnamefont {S.}~\bibnamefont {Henkes}}, \bibinfo {author}
  {\bibfnamefont {B.}~\bibnamefont {Chakraborty}}, \bibinfo {author}
  {\bibfnamefont {C.~S.}\ \bibnamefont {O'Hern}}, \ and\ \bibinfo {author}
  {\bibfnamefont {R.~P.}\ \bibnamefont {Behringer}},\ }\Doi
  {10.1103/PhysRevE.80.060303} {\bibfield  {journal} {\bibinfo  {journal}
  {Physical Review E},\ }\textbf {\bibinfo {volume} {80}},\ \bibinfo {pages}
  {060303} (\bibinfo {year} {2009})}\BibitemShut {NoStop}%
\bibitem [{\citenamefont {Tighe}\ and\ \citenamefont
  {Vlugt}(2011)}]{Tighe2011}%
  \BibitemOpen
  \bibfield  {author} {\bibinfo {author} {\bibfnamefont {B.~P.}\ \bibnamefont
  {Tighe}}\ and\ \bibinfo {author} {\bibfnamefont {T.~J.~H.}\ \bibnamefont
  {Vlugt}},\ }\Doi {10.1088/1742-5468/2011/04/P04002} {\bibfield  {journal}
  {\bibinfo  {journal} {Journal of Statistical Mechanics},\ }\textbf {\bibinfo
  {volume} {2011}},\ \bibinfo {pages} {P04002} (\bibinfo {year}
  {2011})}\BibitemShut {NoStop}%
\bibitem [{\citenamefont {Srebro}\ and\ \citenamefont
  {Levine}(2003)}]{Srebro2003}%
  \BibitemOpen
  \bibfield  {author} {\bibinfo {author} {\bibfnamefont {Y.}~\bibnamefont
  {Srebro}}\ and\ \bibinfo {author} {\bibfnamefont {D.}~\bibnamefont
  {Levine}},\ }\Doi {10.1103/PhysRevE.68.061301} {\bibfield  {journal}
  {\bibinfo  {journal} {Physical Review E},\ }\textbf {\bibinfo {volume}
  {68}},\ \bibinfo {pages} {061301} (\bibinfo {year} {2003})}\BibitemShut
  {NoStop}%
\bibitem [{\citenamefont {Bowles}\ and\ \citenamefont
  {Ashwin}(2011)}]{Bowles2011}%
  \BibitemOpen
  \bibfield  {author} {\bibinfo {author} {\bibfnamefont {R.~K.}\ \bibnamefont
  {Bowles}}\ and\ \bibinfo {author} {\bibfnamefont {S.~S.}\ \bibnamefont
  {Ashwin}},\ }\Doi {10.1103/PhysRevE.83.031302} {\bibfield  {journal}
  {\bibinfo  {journal} {Physical Review E},\ }\textbf {\bibinfo {volume}
  {83}},\ \bibinfo {pages} {031302} (\bibinfo {year} {2011})}\BibitemShut
  {NoStop}%
\bibitem [{\citenamefont {Song}\ \emph {et~al.}(2008)\citenamefont {Song},
  \citenamefont {Wang},\ and\ \citenamefont {Makse}}]{Song2008}%
  \BibitemOpen
  \bibfield  {author} {\bibinfo {author} {\bibfnamefont {C.}~\bibnamefont
  {Song}}, \bibinfo {author} {\bibfnamefont {P.}~\bibnamefont {Wang}}, \ and\
  \bibinfo {author} {\bibfnamefont {H.~A.}\ \bibnamefont {Makse}},\ }\Doi
  {10.1038/nature06981} {\bibfield  {journal} {\bibinfo  {journal} {Nature},\
  }\textbf {\bibinfo {volume} {453}},\ \bibinfo {pages} {629} (\bibinfo {year}
  {2008})}\BibitemShut {NoStop}%
\bibitem [{\citenamefont {Briscoe}\ \emph {et~al.}(2008)\citenamefont
  {Briscoe}, \citenamefont {Song}, \citenamefont {Wang},\ and\ \citenamefont
  {Makse}}]{Briscoe2008}%
  \BibitemOpen
  \bibfield  {author} {\bibinfo {author} {\bibfnamefont {C.}~\bibnamefont
  {Briscoe}}, \bibinfo {author} {\bibfnamefont {C.}~\bibnamefont {Song}},
  \bibinfo {author} {\bibfnamefont {P.}~\bibnamefont {Wang}}, \ and\ \bibinfo
  {author} {\bibfnamefont {H.~A.}\ \bibnamefont {Makse}},\ }\Doi
  {10.1103/PhysRevLett.101.188001} {\bibfield  {journal} {\bibinfo  {journal}
  {Physical Review Letters},\ }\textbf {\bibinfo {volume} {101}},\ \bibinfo
  {pages} {188001} (\bibinfo {year} {2008})}\BibitemShut {NoStop}%
\bibitem [{\citenamefont {Nowak}\ \emph {et~al.}(1998)\citenamefont {Nowak},
  \citenamefont {Knight}, \citenamefont {Ben-Naim}, \citenamefont {Jaeger},\
  and\ \citenamefont {Nagel}}]{Nowak1998}%
  \BibitemOpen
  \bibfield  {author} {\bibinfo {author} {\bibfnamefont {E.~R.}\ \bibnamefont
  {Nowak}}, \bibinfo {author} {\bibfnamefont {J.~B.}\ \bibnamefont {Knight}},
  \bibinfo {author} {\bibfnamefont {E.}~\bibnamefont {Ben-Naim}}, \bibinfo
  {author} {\bibfnamefont {H.~M.}\ \bibnamefont {Jaeger}}, \ and\ \bibinfo
  {author} {\bibfnamefont {S.~R.}\ \bibnamefont {Nagel}},\ }\Doi
  {10.1103/PhysRevE.57.1971} {\bibfield  {journal} {\bibinfo  {journal}
  {Physical Review E},\ }\textbf {\bibinfo {volume} {57}},\ \bibinfo {pages}
  {1971} (\bibinfo {year} {1998})}\BibitemShut {NoStop}%
\bibitem [{\citenamefont {Schr\"{o}ter}\ \emph {et~al.}(2005)\citenamefont
  {Schr\"{o}ter}, \citenamefont {Goldman},\ and\ \citenamefont
  {Swinney}}]{Schroter2005}%
  \BibitemOpen
  \bibfield  {author} {\bibinfo {author} {\bibfnamefont {M.}~\bibnamefont
  {Schr\"{o}ter}}, \bibinfo {author} {\bibfnamefont {D.~I.}\ \bibnamefont
  {Goldman}}, \ and\ \bibinfo {author} {\bibfnamefont {H.~L.}\ \bibnamefont
  {Swinney}},\ }\Doi {10.1103/PhysRevE.71.030301} {\bibfield  {journal}
  {\bibinfo  {journal} {Physical Review E},\ }\textbf {\bibinfo {volume}
  {71}},\ \bibinfo {pages} {030301} (\bibinfo {year} {2005})}\BibitemShut
  {NoStop}%
\bibitem [{\citenamefont {Lechenault}\ \emph {et~al.}(2006)\citenamefont
  {Lechenault}, \citenamefont {Cruz}, \citenamefont {Dauchot},\ and\
  \citenamefont {Bertin}}]{Lechenault2006}%
  \BibitemOpen
  \bibfield  {author} {\bibinfo {author} {\bibfnamefont {F.}~\bibnamefont
  {Lechenault}}, \bibinfo {author} {\bibfnamefont {F.~D.}\ \bibnamefont
  {Cruz}}, \bibinfo {author} {\bibfnamefont {O.}~\bibnamefont {Dauchot}}, \
  and\ \bibinfo {author} {\bibfnamefont {E.}~\bibnamefont {Bertin}},\ }\Doi
  {10.1088/1742-5468/2006/07/P07009} {\bibfield  {journal} {\bibinfo  {journal}
  {Journal of Statistical Mechanics},\ }\textbf {\bibinfo {volume} {2006}},\
  \bibinfo {pages} {P07009} (\bibinfo {year} {2006})}\BibitemShut {NoStop}%
\bibitem [{\citenamefont {Ribi\`{e}re}\ \emph {et~al.}(2007)\citenamefont
  {Ribi\`{e}re}, \citenamefont {Richard}, \citenamefont {Philippe},
  \citenamefont {Bideau},\ and\ \citenamefont {Delannay}}]{Ribiere2007}%
  \BibitemOpen
  \bibfield  {author} {\bibinfo {author} {\bibfnamefont {P.}~\bibnamefont
  {Ribi\`{e}re}}, \bibinfo {author} {\bibfnamefont {P.}~\bibnamefont
  {Richard}}, \bibinfo {author} {\bibfnamefont {P.}~\bibnamefont {Philippe}},
  \bibinfo {author} {\bibfnamefont {D.}~\bibnamefont {Bideau}}, \ and\ \bibinfo
  {author} {\bibfnamefont {R.}~\bibnamefont {Delannay}},\ }\Doi
  {10.1140/epje/e2007-00017-x} {\bibfield  {journal} {\bibinfo  {journal} {The
  European Physical Journal E},\ }\textbf {\bibinfo {volume} {22}},\ \bibinfo
  {pages} {249} (\bibinfo {year} {2007})}\BibitemShut {NoStop}%
\bibitem [{\citenamefont {McNamara}\ \emph {et~al.}(2009)\citenamefont
  {McNamara}, \citenamefont {Richard}, \citenamefont {de~Richter},
  \citenamefont {{Le Ca\"{e}r}},\ and\ \citenamefont
  {Delannay}}]{McNamara2009}%
  \BibitemOpen
  \bibfield  {author} {\bibinfo {author} {\bibfnamefont {S.}~\bibnamefont
  {McNamara}}, \bibinfo {author} {\bibfnamefont {P.}~\bibnamefont {Richard}},
  \bibinfo {author} {\bibfnamefont {S.~K.}\ \bibnamefont {de~Richter}},
  \bibinfo {author} {\bibfnamefont {G.}~\bibnamefont {{Le Ca\"{e}r}}}, \ and\
  \bibinfo {author} {\bibfnamefont {R.}~\bibnamefont {Delannay}},\ }\Doi
  {10.1103/PhysRevE.80.031301} {\bibfield  {journal} {\bibinfo  {journal}
  {Physical Review E},\ }\textbf {\bibinfo {volume} {80}},\ \bibinfo {pages}
  {031301} (\bibinfo {year} {2009})}\BibitemShut {NoStop}%
\bibitem [{\citenamefont {Zhao}\ and\ \citenamefont
  {Schr\"{o}ter}(2012)}]{Zhao2012a}%
  \BibitemOpen
  \bibfield  {author} {\bibinfo {author} {\bibfnamefont {S.-C.}\ \bibnamefont
  {Zhao}}\ and\ \bibinfo {author} {\bibfnamefont {M.}~\bibnamefont
  {Schr\"{o}ter}},\ }\href@noop {} {\bibfield  {journal} {\bibinfo  {journal}
  {in preparation}} (\bibinfo {year} {2012})}\BibitemShut {NoStop}%
\bibitem [{\citenamefont {Majmudar}(2006)}]{Majmudar2006}%
  \BibitemOpen
  \bibfield  {author} {\bibinfo {author} {\bibfnamefont {T.~S.}\ \bibnamefont
  {Majmudar}},\ }\emph {\bibinfo {title} {{Experimental Studies of
  Two-Dimensional Granular Systems Using Grain-Scale Contact Force
  Measurements}}},\ \href@noop {} {Ph.D. thesis},\ \bibinfo  {school} {Duke
  University} (\bibinfo {year} {2006})\BibitemShut {NoStop}%
\bibitem [{\citenamefont {Majmudar}\ \emph {et~al.}(2007)\citenamefont
  {Majmudar}, \citenamefont {Sperl}, \citenamefont {Luding},\ and\
  \citenamefont {Behringer}}]{Majmudar2007}%
  \BibitemOpen
  \bibfield  {author} {\bibinfo {author} {\bibfnamefont {T.~S.}\ \bibnamefont
  {Majmudar}}, \bibinfo {author} {\bibfnamefont {M.}~\bibnamefont {Sperl}},
  \bibinfo {author} {\bibfnamefont {S.}~\bibnamefont {Luding}}, \ and\ \bibinfo
  {author} {\bibfnamefont {R.~P.}\ \bibnamefont {Behringer}},\ }\Doi
  {10.1103/PhysRevLett.98.058001} {\bibfield  {journal} {\bibinfo  {journal}
  {Physical Review Letters},\ }\textbf {\bibinfo {volume} {98}},\ \bibinfo
  {pages} {058001} (\bibinfo {year} {2007})}\BibitemShut {NoStop}%
\bibitem [{\citenamefont {Dean}\ and\ \citenamefont
  {Lef\`{e}vre}(2003)}]{Dean2003}%
  \BibitemOpen
  \bibfield  {author} {\bibinfo {author} {\bibfnamefont {D.~S.}\ \bibnamefont
  {Dean}}\ and\ \bibinfo {author} {\bibfnamefont {A.}~\bibnamefont
  {Lef\`{e}vre}},\ }\Doi {10.1103/PhysRevLett.90.198301} {\bibfield  {journal}
  {\bibinfo  {journal} {Physical Review Letters},\ }\textbf {\bibinfo {volume}
  {90}},\ \bibinfo {pages} {198301} (\bibinfo {year} {2003})}\BibitemShut
  {NoStop}%
\bibitem [{\citenamefont {Puckett}()}]{peDiscSolve}%
  \BibitemOpen
  \bibfield  {author} {\bibinfo {author} {\bibfnamefont {J.~G.}\ \bibnamefont
  {Puckett}},\ }\href {http://nile.physics.ncsu.edu/pub/peDiscSolve/}
  {}\bibinfo {note} {{Photoelastic disc solver: {\tt
  http://nile.physics.ncsu.edu/pub/peDiscSolve/}}}\BibitemShut {NoStop}%
\bibitem [{\citenamefont {Howell}\ \emph {et~al.}(1999)\citenamefont {Howell},
  \citenamefont {Behringer},\ and\ \citenamefont {Veje}}]{Howell1999}%
  \BibitemOpen
  \bibfield  {author} {\bibinfo {author} {\bibfnamefont {D.}~\bibnamefont
  {Howell}}, \bibinfo {author} {\bibfnamefont {R.~P.}\ \bibnamefont
  {Behringer}}, \ and\ \bibinfo {author} {\bibfnamefont {C.}~\bibnamefont
  {Veje}},\ }\Doi {10.1103/PhysRevLett.82.5241} {\bibfield  {journal} {\bibinfo
   {journal} {Physical Review Letters},\ }\textbf {\bibinfo {volume} {82}},\
  \bibinfo {pages} {5241} (\bibinfo {year} {1999})}\BibitemShut {NoStop}%
\bibitem [{\citenamefont {O'Hern}\ \emph {et~al.}(2003)\citenamefont {O'Hern},
  \citenamefont {Silbert}, \citenamefont {Liu},\ and\ \citenamefont
  {Nagel}}]{OHern2003}%
  \BibitemOpen
  \bibfield  {author} {\bibinfo {author} {\bibfnamefont {C.~S.}\ \bibnamefont
  {O'Hern}}, \bibinfo {author} {\bibfnamefont {L.~E.}\ \bibnamefont {Silbert}},
  \bibinfo {author} {\bibfnamefont {A.~J.}\ \bibnamefont {Liu}}, \ and\
  \bibinfo {author} {\bibfnamefont {S.~R.}\ \bibnamefont {Nagel}},\ }\Doi
  {10.1103/PhysRevE.68.011306} {\bibfield  {journal} {\bibinfo  {journal}
  {Physical Review E},\ }\textbf {\bibinfo {volume} {68}},\ \bibinfo {pages}
  {011306} (\bibinfo {year} {2003})}\BibitemShut {NoStop}%
\bibitem [{\citenamefont {Shen}\ \emph {et~al.}(2012)\citenamefont {Shen},
  \citenamefont {O'Hern},\ and\ \citenamefont {Shattuck}}]{Shen2012b}%
  \BibitemOpen
  \bibfield  {author} {\bibinfo {author} {\bibfnamefont {T.}~\bibnamefont
  {Shen}}, \bibinfo {author} {\bibfnamefont {C.~S.}\ \bibnamefont {O'Hern}}, \
  and\ \bibinfo {author} {\bibfnamefont {M.~D.}\ \bibnamefont {Shattuck}},\
  }\Doi {10.1103/PhysRevE.85.011308} {\bibfield  {journal} {\bibinfo  {journal}
  {Physical Review E},\ }\textbf {\bibinfo {volume} {85}},\ \bibinfo {pages}
  {011308} (\bibinfo {year} {2012})}\BibitemShut {NoStop}%
\bibitem [{\citenamefont {Rycroft}\ \emph {et~al.}(2006)\citenamefont
  {Rycroft}, \citenamefont {Grest}, \citenamefont {Landry},\ and\ \citenamefont
  {Bazant}}]{Rycroft2006}%
  \BibitemOpen
  \bibfield  {author} {\bibinfo {author} {\bibfnamefont {C.~H.}\ \bibnamefont
  {Rycroft}}, \bibinfo {author} {\bibfnamefont {G.~S.}\ \bibnamefont {Grest}},
  \bibinfo {author} {\bibfnamefont {J.~W.}\ \bibnamefont {Landry}}, \ and\
  \bibinfo {author} {\bibfnamefont {M.~Z.}\ \bibnamefont {Bazant}},\ }\Doi
  {10.1103/PhysRevE.74.021306} {\bibfield  {journal} {\bibinfo  {journal}
  {Physical Review E},\ }\textbf {\bibinfo {volume} {74}},\ \bibinfo {pages}
  {021306} (\bibinfo {year} {2006})}\BibitemShut {NoStop}%
\bibitem [{\citenamefont {Mueth}\ \emph {et~al.}(1998)\citenamefont {Mueth},
  \citenamefont {Jaeger},\ and\ \citenamefont {Nagel}}]{Mueth1998}%
  \BibitemOpen
  \bibfield  {author} {\bibinfo {author} {\bibfnamefont {D.~M.}\ \bibnamefont
  {Mueth}}, \bibinfo {author} {\bibfnamefont {H.~M.}\ \bibnamefont {Jaeger}}, \
  and\ \bibinfo {author} {\bibfnamefont {S.~R.}\ \bibnamefont {Nagel}},\ }\Doi
  {10.1103/PhysRevE.57.3164} {\bibfield  {journal} {\bibinfo  {journal}
  {Physical Review E},\ }\textbf {\bibinfo {volume} {57}},\ \bibinfo {pages}
  {3164} (\bibinfo {year} {1998})}\BibitemShut {NoStop}%
\bibitem [{\citenamefont {L{\o}voll}\ \emph {et~al.}(1999)\citenamefont
  {L{\o}voll}, \citenamefont {M{\aa}l{\o}y},\ and\ \citenamefont
  {Flekk{\o}y}}]{Lovoll1999}%
  \BibitemOpen
  \bibfield  {author} {\bibinfo {author} {\bibfnamefont {G.}~\bibnamefont
  {L{\o}voll}}, \bibinfo {author} {\bibfnamefont {K.~J.}\ \bibnamefont
  {M{\aa}l{\o}y}}, \ and\ \bibinfo {author} {\bibfnamefont {E.~G.}\
  \bibnamefont {Flekk{\o}y}},\ }\Doi {10.1103/PhysRevE.60.5872} {\bibfield
  {journal} {\bibinfo  {journal} {Physical Review E},\ }\textbf {\bibinfo
  {volume} {60}},\ \bibinfo {pages} {5872} (\bibinfo {year}
  {1999})}\BibitemShut {NoStop}%
\bibitem [{\citenamefont {Blair}\ \emph {et~al.}(2001)\citenamefont {Blair},
  \citenamefont {Mueggenburg}, \citenamefont {Marshall}, \citenamefont
  {Jaeger},\ and\ \citenamefont {Nagel}}]{Blair2001}%
  \BibitemOpen
  \bibfield  {author} {\bibinfo {author} {\bibfnamefont {D.~L.}\ \bibnamefont
  {Blair}}, \bibinfo {author} {\bibfnamefont {N.~W.}\ \bibnamefont
  {Mueggenburg}}, \bibinfo {author} {\bibfnamefont {A.~H.}\ \bibnamefont
  {Marshall}}, \bibinfo {author} {\bibfnamefont {H.~M.}\ \bibnamefont
  {Jaeger}}, \ and\ \bibinfo {author} {\bibfnamefont {S.~R.}\ \bibnamefont
  {Nagel}},\ }\Doi {10.1103/PhysRevE.63.041304} {\bibfield  {journal} {\bibinfo
   {journal} {Physical Review E},\ }\textbf {\bibinfo {volume} {63}},\ \bibinfo
  {pages} {041304} (\bibinfo {year} {2001})}\BibitemShut {NoStop}%
\bibitem [{\citenamefont {Makse}\ \emph {et~al.}(2000)\citenamefont {Makse},
  \citenamefont {Johnson},\ and\ \citenamefont {Schwartz}}]{Makse2000}%
  \BibitemOpen
  \bibfield  {author} {\bibinfo {author} {\bibfnamefont {H.~A.}\ \bibnamefont
  {Makse}}, \bibinfo {author} {\bibfnamefont {D.~L.}\ \bibnamefont {Johnson}},
  \ and\ \bibinfo {author} {\bibfnamefont {L.~M.}\ \bibnamefont {Schwartz}},\
  }\href {http://www.ncbi.nlm.nih.gov/pubmed/10990635} {\bibfield  {journal}
  {\bibinfo  {journal} {Physical Review Letters},\ }\textbf {\bibinfo {volume}
  {84}},\ \bibinfo {pages} {4160} (\bibinfo {year} {2000})}\BibitemShut
  {NoStop}%
\bibitem [{\citenamefont {Corwin}\ \emph {et~al.}(2005)\citenamefont {Corwin},
  \citenamefont {Jaeger},\ and\ \citenamefont {Nagel}}]{Corwin2005}%
  \BibitemOpen
  \bibfield  {author} {\bibinfo {author} {\bibfnamefont {E.~I.}\ \bibnamefont
  {Corwin}}, \bibinfo {author} {\bibfnamefont {H.~M.}\ \bibnamefont {Jaeger}},
  \ and\ \bibinfo {author} {\bibfnamefont {S.~R.}\ \bibnamefont {Nagel}},\
  }\Doi {10.1038/nature03698} {\bibfield  {journal} {\bibinfo  {journal}
  {Nature},\ }\textbf {\bibinfo {volume} {435}},\ \bibinfo {pages} {1075}
  (\bibinfo {year} {2005})}\BibitemShut {NoStop}%
\bibitem [{\citenamefont {Mukhopadhyay}\ and\ \citenamefont
  {Peixinho}(2011)}]{Mukhopadhyay2011}%
  \BibitemOpen
  \bibfield  {author} {\bibinfo {author} {\bibfnamefont {S.}~\bibnamefont
  {Mukhopadhyay}}\ and\ \bibinfo {author} {\bibfnamefont {J.}~\bibnamefont
  {Peixinho}},\ }\Doi {10.1103/PhysRevE.84.011302} {\bibfield  {journal}
  {\bibinfo  {journal} {Physical Review E},\ }\textbf {\bibinfo {volume}
  {84}},\ \bibinfo {pages} {011302} (\bibinfo {year} {2011})}\BibitemShut
  {NoStop}%
\bibitem [{\citenamefont {Saadatfar}\ \emph {et~al.}(2012)\citenamefont
  {Saadatfar}, \citenamefont {Sheppard}, \citenamefont {Senden},\ and\
  \citenamefont {Kabla}}]{Saadatfar2012}%
  \BibitemOpen
  \bibfield  {author} {\bibinfo {author} {\bibfnamefont {M.}~\bibnamefont
  {Saadatfar}}, \bibinfo {author} {\bibfnamefont {A.~P.}\ \bibnamefont
  {Sheppard}}, \bibinfo {author} {\bibfnamefont {T.~J.}\ \bibnamefont
  {Senden}}, \ and\ \bibinfo {author} {\bibfnamefont {A.~J.}\ \bibnamefont
  {Kabla}},\ }\Doi {10.1016/j.jmps.2011.10.001} {\bibfield  {journal} {\bibinfo
   {journal} {Journal of the Mechanics and Physics of Solids},\ }\textbf
  {\bibinfo {volume} {60}},\ \bibinfo {pages} {55} (\bibinfo {year}
  {2012})}\BibitemShut {NoStop}%
\bibitem [{\citenamefont {Bruji\'{c}}\ \emph {et~al.}(2003)\citenamefont
  {Bruji\'{c}}, \citenamefont {Edwards}, \citenamefont {Grinev}, \citenamefont
  {Hopkinson}, \citenamefont {Bruji\'{c}},\ and\ \citenamefont
  {Makse}}]{Brujic2003}%
  \BibitemOpen
  \bibfield  {author} {\bibinfo {author} {\bibfnamefont {J.}~\bibnamefont
  {Bruji\'{c}}}, \bibinfo {author} {\bibfnamefont {S.~F.}\ \bibnamefont
  {Edwards}}, \bibinfo {author} {\bibfnamefont {D.~V.}\ \bibnamefont {Grinev}},
  \bibinfo {author} {\bibfnamefont {I.}~\bibnamefont {Hopkinson}}, \bibinfo
  {author} {\bibfnamefont {D.}~\bibnamefont {Bruji\'{c}}}, \ and\ \bibinfo
  {author} {\bibfnamefont {H.~A.}\ \bibnamefont {Makse}},\ }\Doi
  {10.1039/b204414e} {\bibfield  {journal} {\bibinfo  {journal} {Faraday
  Discussions},\ }\textbf {\bibinfo {volume} {123}},\ \bibinfo {pages} {207}
  (\bibinfo {year} {2003})}\BibitemShut {NoStop}%
\bibitem [{\citenamefont {Zhou}\ \emph {et~al.}(2006)\citenamefont {Zhou},
  \citenamefont {Long}, \citenamefont {Wang},\ and\ \citenamefont
  {Dinsmore}}]{Zhou2006a}%
  \BibitemOpen
  \bibfield  {author} {\bibinfo {author} {\bibfnamefont {J.}~\bibnamefont
  {Zhou}}, \bibinfo {author} {\bibfnamefont {S.}~\bibnamefont {Long}}, \bibinfo
  {author} {\bibfnamefont {Q.}~\bibnamefont {Wang}}, \ and\ \bibinfo {author}
  {\bibfnamefont {A.~D.}\ \bibnamefont {Dinsmore}},\ }\Doi
  {10.1126/science.1125151} {\bibfield  {journal} {\bibinfo  {journal}
  {Science},\ }\textbf {\bibinfo {volume} {312}},\ \bibinfo {pages} {1631}
  (\bibinfo {year} {2006})}\BibitemShut {NoStop}%
\bibitem [{\citenamefont {Desmond}\ \emph {et~al.}(2012)\citenamefont
  {Desmond}, \citenamefont {Young}, \citenamefont {Chen},\ and\ \citenamefont
  {Weeks}}]{Desmond2012}%
  \BibitemOpen
  \bibfield  {author} {\bibinfo {author} {\bibfnamefont {K.~W.}\ \bibnamefont
  {Desmond}}, \bibinfo {author} {\bibfnamefont {P.~J.}\ \bibnamefont {Young}},
  \bibinfo {author} {\bibfnamefont {D.}~\bibnamefont {Chen}}, \ and\ \bibinfo
  {author} {\bibfnamefont {E.~R.}\ \bibnamefont {Weeks}},\ }\href
  {http://arxiv.org/abs/1206.0070} { (\bibinfo {year} {2012})},\ \Eprint
  {http://arxiv.org/abs/arxiv.org: 1206.0070} {arXiv:arxiv.org: 1206.0070}
  \BibitemShut {NoStop}%
\bibitem [{\citenamefont {Wang}\ \emph {et~al.}(2012)\citenamefont {Wang},
  \citenamefont {Song}, \citenamefont {Wang},\ and\ \citenamefont
  {Makse}}]{Wang2012d}%
  \BibitemOpen
  \bibfield  {author} {\bibinfo {author} {\bibfnamefont {K.}~\bibnamefont
  {Wang}}, \bibinfo {author} {\bibfnamefont {C.}~\bibnamefont {Song}}, \bibinfo
  {author} {\bibfnamefont {P.}~\bibnamefont {Wang}}, \ and\ \bibinfo {author}
  {\bibfnamefont {H.~A.}\ \bibnamefont {Makse}},\ }\Doi
  {10.1103/PhysRevE.86.011305} {\bibfield  {journal} {\bibinfo  {journal}
  {Physical Review E},\ }\textbf {\bibinfo {volume} {86}},\ \bibinfo {pages}
  {011305} (\bibinfo {year} {2012})}\BibitemShut {NoStop}%
\bibitem [{\citenamefont {Blumenfeld}\ \emph {et~al.}(2012)\citenamefont
  {Blumenfeld}, \citenamefont {Jordan},\ and\ \citenamefont
  {Edwards}}]{Blumenfeld2012}%
  \BibitemOpen
  \bibfield  {author} {\bibinfo {author} {\bibfnamefont {R.}~\bibnamefont
  {Blumenfeld}}, \bibinfo {author} {\bibfnamefont {J.~F.}\ \bibnamefont
  {Jordan}}, \ and\ \bibinfo {author} {\bibfnamefont {S.~F.}\ \bibnamefont
  {Edwards}},\ }\href {http://arxiv.org/abs/1204.2977} { (\bibinfo {year}
  {2012})},\ \Eprint {http://arxiv.org/abs/arxiv.org: 1204.2977} {arxiv.org:
  1204.2977} \BibitemShut {NoStop}%
\bibitem [{\citenamefont {Reynolds}(1885)}]{Reynolds1885}%
  \BibitemOpen
  \bibfield  {author} {\bibinfo {author} {\bibfnamefont {O.}~\bibnamefont
  {Reynolds}},\ }\href@noop {} {\bibfield  {journal} {\bibinfo  {journal}
  {Philos. Mag. Ser. 5},\ }\textbf {\bibinfo {volume} {20}},\ \bibinfo {pages}
  {469} (\bibinfo {year} {1885})}\BibitemShut {NoStop}%
\end{thebibliography}

\end{document}